\documentclass{article}

\usepackage{arxiv}
\usepackage{makecell}
\usepackage[utf8]{inputenc} 
\usepackage[T1]{fontenc}    
\usepackage{hyperref}       
\usepackage{color}
\usepackage[figuresright]{rotating}
\usepackage{url}            
\usepackage{booktabs}       
\usepackage{amsfonts}       
\usepackage{nicefrac}       
\usepackage{microtype}      
\usepackage{lipsum,amsmath}
\usepackage{graphicx}
\usepackage{natbib}
\usepackage{algpseudocode,algorithm,algorithmicx}
\setcitestyle{authoryear,open={(},close={)}}

\usepackage{setspace}
\title{BayesSRW: Bayesian Sampling and Re-weighting approach for variance reduction}

\author{
  Carol Liu \\
  Department of Marketing\\
  University of Wisconsin-Madison\\
  Madison, WI, USA\\
  \texttt{liu747@wisc.edu} \\
}

\begin{document}
\maketitle

\begin{abstract}

In this paper, we address the challenge of sampling in scenarios where limited resources prevent exhaustive measurement across all subjects. We consider a setting where samples are drawn from multiple groups, each following a distribution with unknown mean and variance parameters. We introduce a novel sampling strategy, motivated simply by Cauchy-Schwarz inequality, which minimizes the variance of the population mean estimator by allocating samples proportionally to both the group size and the standard deviation. This approach improves the efficiency of sampling by focusing resources on groups with greater variability, thereby enhancing the precision of the overall estimate. Additionally, we extend our method to a two-stage sampling procedure in a Bayes approach, named \verb+BayesSRW+, where a preliminary stage is used to estimate the variance, which then informs the optimal allocation of the remaining sampling budget. Through simulation examples, we demonstrate the effectiveness of our approach in reducing estimation uncertainty and providing more reliable insights in applications ranging from user experience surveys to high-dimensional peptide array studies.

\end{abstract}


\section{Introduction and problem set up}

Suppose we have $N$ samples from $k$ groups, each with sample size $N_i$ ($\sum_{i=1}^{k}N_i = N$). Within each group, $N_i$ samples follow an iid normal distribution with mean parameter $\mu_i$ and variance parameter $\sigma_i^2$ . 
Specifically, 
\begin{eqnarray*} X_{i}^{j} \sim {\rm N}(\mu_i, \sigma_i^2), \ i\in {1,2,\dots, k}, \ j \in {1,2,\dots,N_i}. \end{eqnarray*} 
Note that the normal distribution can be easily generated, and the entire discussion can hold true with the mean parameter $\mu_i$ and variance parameter $\sigma_i^2$ without the specific assumptions or limitations on distribution. We set this in a parametric way for simplicity. There are numerous practical applications. For example, Google uses the Google Opinion Rewards program (\href{https://surveys.google.com/google-opinion-rewards/}{GOR}) extensively to evaluate user experience (e.g., on Google search) by assigning surveys to registered users. Google Opinion Rewards is a platform where users receive surveys in exchange for rewards, typically in the form of Google Play credits. These surveys are designed to capture user feedback on various products and services, allowing Google to gather valuable insights into user preferences, satisfaction, and behavior. The surveys are distributed to a diverse and global user base, providing a broad spectrum of opinions across different demographics and regions. Each survey response can be considered as an individual data point ($X_i^j$), where $i$ represents different user segments or query types, and $j$ corresponds to specific responses within those segments. The flexibility of the GOR platform allows for the rapid deployment of surveys, enabling Google to test and refine new features or changes before a wider rollout. This is particularly valuable in assessing the potential impact of changes to core services like Google Search. The program's design also allows for targeted surveys, ensuring that the feedback is relevant to the users' experience. However, while the GOR program provides a scalable way to gather user feedback, it is still not feasible to survey every eligible user due to cost constraints and the potential risk of survey fatigue, which could negatively impact the user experience. In this setting, $X_i^j$ can represent a survey response on each query, and $k$ can represent the query or user segmentation. $\mu_i$ is the treatment effect on that query, and $\sigma_i$ measures uncertainty. Survey-based research is a prime example of lab experiments, heavily applied in social sciences to explore human behavior, cognition, and decision-making processes. For instance, researchers used a controlled experimental setup to evaluate the efficacy of different retrieval practices on memory enhancement \citep{liu2022progressive}. This research exemplifies how well-designed surveys and lab experiments can uncover significant insights into cognitive processes, similar to how the GOR platform gathers user experience data for product refinement.

Similarly, in the high-density peptide array problem, $X_i^j$ measures the fold difference (or treatment effect) between groups on a specific peptide, and $k$ represents cluster information, such as protein groups or domains. High-density peptide arrays are a powerful tool in proteomics, enabling the simultaneous measurement of interactions, modifications, or expression levels across a vast number of peptides. These arrays are particularly useful in identifying disease-specific biomarkers or understanding the underlying mechanisms of diseases. For example, in a recent study, the fold difference between control subjects and those with Rheumatoid Arthritis (RA) was evaluated across $N = 172,828$ peptides \citep{zheng2020disordered,mergaert2022rheumatoid}. This comprehensive analysis allowed the identification of peptides that were significantly associated with RA, providing insights into potential diagnostic markers and therapeutic targets. Furthermore, in even more extensive studies, such as those discussed by \citep{parker2024novel} and \citep{amjadi2024novel}, the scope of analysis extended to the entire human proteome. These studies examined the fold difference across approximately $N = 5.3$ million peptides, representing nearly the entirety of the human proteome. 

Although identifying the signal at each subject level ($X_i^j$) is valuable and can provide granular insights, many investigators are often more interested in the population mean across all groups or supports, which offers a more comprehensive view of the treatment effect. The population mean is calculated as follows:
\begin{eqnarray*} \mu = \frac{1}{N}\sum_{i=1}^k N_i \mu_i. \end{eqnarray*}
In the context of the Google Opinion Rewards (GOR) example discussed earlier, this population mean represents the overall treatment effect on all Google users. This aggregate metric is often more relevant and scalable compared to evaluating the treatment effect on a specific subset of users or a particular segment of queries. By focusing on the population mean, researchers and practitioners can derive conclusions that are generalizable and applicable to the broader user base, making the findings more actionable for large-scale interventions. For instance, while understanding the treatment effect within a specific query segment might help fine-tune the user experience for that segment, knowing the overall impact across all users allows for more informed decisions when deploying changes at scale. This approach aligns with the broader objectives in fields such as marketing, where the goal is often to optimize the overall effectiveness of a campaign or product feature rather than just a localized improvement. Similarly, in other domains like clinical trials or social sciences, researchers often aim to generalize findings to the entire population of interest, ensuring that the interventions are effective on a wide scale.

However, in many practical scenarios, it is not feasible to measure each individual sample ($X_i^j$). For instance, assigning surveys to all eligible users would require substantial budgetary resources and could pose significant risks to user experience, particularly if the treatment is still in an early experimental phase. Such widespread interventions could negatively impact user satisfaction and engagement, especially if the changes are not yet fully refined. As a result, a more selective approach is often necessary. Similarly, in the high-density peptide array problem, while it is theoretically possible to measure the treatment effect on each peptide across the entire human proteome, this would demand considerable effort and funding. The scale of such an undertaking, involving millions of peptides, would require extensive processing, careful handling, and detailed analysis. This process is not only resource-intensive but also time-consuming, making it impractical for routine applications. Consequently, researchers often focus on specific subsets or clusters of peptides that are most likely to yield significant results, based on prior knowledge or preliminary studies. While this approach is more manageable, it does involve trade-offs in the comprehensiveness and granularity of the findings. Typically, there is only a budget for $n$ observations where $n < N$, leading to the critical question: how should these $n$ observations be optimally assigned for measurement? In the discussion below, we will use the survey scenario as an example. 

\section{Methodology and algorithm}

\subsection{A simple idea from Cauchy inequality}

A naive approach to survey analysis might involve uniformly distributing surveys across all $N$ queries and estimating the overall effect as the arithmetic mean of the $n$ selected queries. While this method provides an unbiased estimate, it is suboptimal when considering variance. Uniform allocation does not account for the potential variability in responses across different queries, which can lead to higher overall variance in the estimate. In practical scenarios, minimizing variance is often more critical and challenging than achieving unbiased estimates alone. The uniform approach treats all queries equally, neglecting the fact that some queries may exhibit greater variability or provide more informative data. For example, certain user segments or query types might show a wider range of responses due to underlying differences in user behavior or preferences. Ignoring this variability can result in inefficient estimates with higher uncertainty. Effective statistical analysis requires strategies that account for response variability to improve precision and reliability of estimates \citep{gelman2007data, richard2007efficient, liu2009comparative, kalton2020introduction}.

The general framework for addressing this problem involves solving the optimization problem of finding the optimal $n_i$, where $i = 1, 2, \ldots, k$, subject to the constraint $\sum_i n_i = n$, such that the associated estimator is unbiased and has minimized variance. In the uniformly sampling scenario, $n_i$ is chosen proportional to $N_i$. Specifically, this approach ensures that $n_i/N_i = \lambda$ for all $i$, where $\lambda$ can be calculated as $\lambda = n/N$. To guarantee an unbiased estimation of $\mu$, inverse probability weighting (IPW) is required. For each group $i$, the sampling weight is $w_i = n_i/N_i$, and the estimator is given by:
\begin{eqnarray*} \hat{\mu} = \frac{\sum_{i=1}^{k} \sum_{j=1}^{n_i} \frac{N_i}{n_i} X_{i}^j}{\sum_{i=1}^{k} \sum_{j=1}^{n_i} \frac{N_i}{n_i}} \end{eqnarray*}

Here, $X_{i}^j$ represents the $j$-th observation from group $i$. This estimator ensures that each group is weighted according to its proportion in the population, thereby achieving an unbiased estimate of the population mean $\mu$. However, while this approach provides an unbiased estimate, it may not always be optimal in terms of variance. To minimize variance, one must consider additional strategies that account for the heterogeneity of responses across different groups, potentially leading to more efficient sampling designs that reduce overall estimation uncertainty. Again, the uniform sampling scenario is just one degenerating case where $N_i/n_i$ is a constant.

It is easy to examine $\hat \mu$ is unbiased for whatever $\{n_i\}$ assignment.
\begin{eqnarray*}
\mathbb{E}(\hat \mu) = \frac{\sum_{i}^{k} \sum_{j}^{n_i} \frac{N_i}{n_i} \mu_i} {\sum_{i}^{k} \sum_{j}^{n_i} \frac{N_i}{n_i}} = \frac{\sum_{i}^{k} N_i \mu_i}{N} = \mu
\end{eqnarray*}

Now we calculate its variance:
\begin{eqnarray*}
\mathbb{V}(\hat \mu) = \frac{\sum_i^k \sum_j^{n_i} \left(\frac{N_i}{n_i}\right)^2 \sigma_i^2}{\left(\sum_i^k \sum_j^{n_i} \frac{Ni}{n_i}\right)^2} = \frac{1}{N^2} \sum_i^k \sum_j^{n_i} \left(\frac{N_i}{n_i}\right)^2 \sigma_i^2 = \frac{1}{N^2} \sum_i^k \frac{N_i^2 \sigma_i^2}{n_i} \geq \frac{1}{N^2} \frac{1}{n} \left(\sum_{i}^{k}N_i \sigma_i\right)^2
\end{eqnarray*}
where the inequality suffices from Cauchy inequality and the equality holds when $\frac{N_i \sigma_i}{n_i}$ is a constant. Therefore we have the following $n_i$ assignment:
\begin{eqnarray*}
n_i^{*} = n \frac{N_i \sigma_i}{\sum_i N_i \sigma_i}
\end{eqnarray*}
Besides the simple mathematical guarantees from Cauchy inequality, the interpretation of the optimal assignment in survey sampling becomes clearer when we shift from assigning surveys proportional to $N_i$ to assigning them proportional to $N_i \sigma_i$, where $\sigma_i$ represents the standard deviation or uncertainty within each group. This optimal assignment strategy ensures that the allocation of surveys accounts not only for the size of each group but also for the level of variability or uncertainty in their responses. Specifically, this approach is invariant with respect to the location parameter $\mu_i$, which denotes the mean of the group, but it heavily relies on the knowledge of the uncertainty represented by $\sigma_i$. Intuitively, this means that more resources or traffic should be allocated to cohorts where there is greater uncertainty (i.e., larger $\sigma_i$). This approach maximizes the efficiency of the survey process by focusing efforts on groups where we have less confidence in the estimates. Conversely, for cohorts where we have higher confidence (i.e., smaller $\sigma_i$), fewer samples are needed as they can adequately represent the distribution with relatively low variance. By employing inverse probability weighting (IPW), we adjust for the varying probabilities of selection and ensure that the estimator remains unbiased. Overall, this strategy allows for a more nuanced and efficient allocation of resources, improving the precision of estimates and reducing overall estimation uncertainty \citep{kim2010calibration}.

\subsection{Two-stage survey sampling and re-weighting}

The previous discussion highlights that optimal survey sampling should allocate the number of observations proportionally to $N_i \sigma_i$, rather than just $N_i$, to better account for variance uncertainty. This approach improves estimation efficiency by focusing resources on groups with higher variability, thus reducing overall estimation variance. However, in practical applications, while $N_i$ is typically known, the variance parameter $\sigma_i$ is often not available as prior knowledge. For instance, in the GOR framework, where each group $i$ could represent a query space (e.g., categorized by product vertical) or a user segment (e.g., demographic slices), obtaining accurate estimates of variance distribution can be challenging.

In such scenarios, a practical solution is to allocate a portion of the sampling budget specifically for variance estimation. This can be implemented through a two-stage sampling and re-weighting procedure. In the first stage, a proportion $p$ of the total sample size $n$ is used to estimate the variance $\sigma_i$ for each group $i$. This preliminary stage helps in assessing the variability across different groups. In the second stage, the remaining portion of the sampling budget is allocated based on the updated estimates of $N_i \sigma_i$, to estimate $\mu$.

\begin{algorithm}[H]
\label{TwoStageSamplingAndReweighting}
  \caption{Two-stage sampling and re-weighting}
  \hspace*{0.02in} 
  {\bf Input:}  Total sample size $N$, sampling budget size $n$, sampling proportion for variance knowledge $p$, group size $k$, $N_i$ for each $i$,  and observations to be measured $X$ \\
  {\bf Output:} $\hat \mu$
  \begin{algorithmic}[1]
  \State Calculate $np$, the number of observations used for variance knowledge.
  \State Do uniform sampling across $k$ groups, where size of each group being $np \frac{N_i}{N}$.
  \State Estimate $\hat \sigma_i$ using observations $np \frac{N_i}{N}$ for each group.
  \State Do optimal sampling across $k$ groups,  where size of each group being $n(1-p) \frac{N_i \hat \sigma_i}{\sum_i N_i \hat \sigma_i}$.
  \State Estimate $\hat \mu_i$ using observations and get $\hat \mu$ following IPW rule. 
  \end{algorithmic}
\end{algorithm}

The proposed algorithm is practically feasible, although it involves one hyperparameter, $p$, which represents the proportion of the total sampling budget allocated for variance estimation. A higher value of $p$ allows for more precise variance estimation but results in a smaller remaining budget of $(1-p)n$ for estimating the parameter of interest. This creates a trade-off between the precision of the variance estimates and the resources available for obtaining accurate estimates of the parameter. Despite this theoretical trade-off, we will provide a robust practical guideline for selecting $p$ in the next section, which will help in balancing these competing considerations effectively.

\subsection{BayesSRW: Bayesian sampling and re-weighting}

As discussed, the first stage of the proposed algorithm involves using the initial sampling to estimate $\sigma_i$ for optimal allocation. However, a Bayesian approach can leverage additional information to enhance this process. In this section, we introduce our algorithm, \verb+BayesSRW+, which extends the two-stage sampling and re-weighting framework by incorporating Bayesian methods. In \verb+BayesSRW+, the first stage estimates the prior distribution for the variance parameters $\sigma_i$ using the initial sample. This prior distribution incorporates prior knowledge or beliefs about the variability in each group. In the second stage, the remaining sampling budget is used to update this prior distribution based on the observed data, refining our estimates of $\sigma_i$ and consequently improving the overall allocation strategy. The final estimate produced by \verb+BayesSRW+ is derived from Bayes' theorem, which integrates the prior distribution with the likelihood of the observed data. This approach effectively leads to a Maximum A Posteriori (MAP) estimator, which provides a balance between prior beliefs and observed evidence. By using Bayesian inference, \verb+BayesSRW+ not only refines the variance estimates but also incorporates uncertainty in a principled manner, potentially offering more robust and accurate parameter estimates compared to traditional methods.

The Bayesian approach is widely recognized for its ability to make robust estimations by integrating prior knowledge with observed data. This methodology is particularly valuable in scenarios where prior information about parameters or variance is available and can be systematically incorporated into the estimation process \citep{benjamini1995controlling, seltzer1996bayesian, efron2001empirical, storey2002direct,bernardo2007bayesian, efron2007correlation, stephens2017false, narisetty2020bayesian, zheng2024bootstrap}. By combining prior beliefs with empirical evidence, Bayesian methods can produce more reliable estimates, especially when dealing with limited or noisy data. In our specific example, we are interested in estimating the joint distribution $p(\mu, \sigma)$. Although estimating probabilities in multiple dimensions can complicate the problem, assuming independence between the location parameter and the scale parameter—such that $p(\mu, \sigma) = g(\mu)h(\sigma)$—is often a practical and reasonable approach \citep{zheng2021mixtwice}. This assumption simplifies the estimation process while still providing valuable insights into the parameter distributions. Additionally, while Inverse Probability Weighting (IPW) is not explicitly cited in \verb+BayesSRW+, it is important to note that employing a weighted approach, proportional to $n_i$, is essential for ensuring that the final estimation is unbiased with respect to the population distribution.

\begin{algorithm}[H]
\label{BayesSRW}
  \caption{BayesSRW: Bayesian sampling and re-weighting}
  \hspace*{0.02in} 
  {\bf Input:}  Total sample size $N$, sampling budget size $n$, sampling proportion for variance knowledge $p$, group size $k$, $N_i$ for each $i$,  and observations to be measured $X$ \\
  {\bf Output:} $\hat \mu$
  \begin{algorithmic}[1]
  \State Calculate $np$, the number of observations used for variance knowledge.
  \State Do uniform sampling across $k$ groups, where size of each group being $np \frac{N_i}{N}$.
  \State Estimate $(\hat \mu_i ^{(1)}, \hat \sigma_i ^ {(1)})$ using observations $np \frac{N_i}{N}$ for each group.
  \State Estimate the distribution of $p(\mu, \sigma)$ from the estimations.
  \State Do optimal sampling across $k$ groups,  where size of each group being $n(1-p) \frac{N_i \hat \sigma_i}{\sum_i N_i \hat \sigma_i}$.
  \State Estimate $(\hat \mu_i ^{(2)}, \hat \sigma_i ^ {(2)})$ using observations $np \frac{N_i}{N}$ for each group and update $p(\mu, \sigma)$.
  \State Get $\hat \mu$ from $p(\mu, \sigma)$. 
  \end{algorithmic}
\end{algorithm}

\section{Simulation and Examples}

There are multiple questions that we need to better understand the application on \verb+BayesSRW+. The first one starts with the question: if we do \verb+BayesSRW+ rather than uniform sampling, how much reduction on confidence interval can we get? 

\begin{itemize}
    \item The variance of random sampling is proportional to $V_0 = \left(\sum_i^k N_i \sigma_i^2 \right)\left(\sum_i N_i\right)$
    \item The optimal variance is proportional to $V_* = \left( \sum_i^k N_i\sigma_i \right)^2$
\end{itemize}

For a simplest scenario, let's start with a $k=2$ scenario. The relative difference for variance reduction, under $k=2$ is: $$\frac{V_*}{V_0} = \frac{\left(N_1 \sigma_1 + N_2 \sigma_2\right)^2}{(N_1\sigma_1^2 + N_2\sigma_2^2)(N_1 + N_2)}$$

With equal sample size $N_1 = N_2$ and $t = \sigma_2/\sigma_1$, $$\frac{V_*}{V_0} = \frac{(1+t)^2}{2(1+t^2)} := f(t)$$

This $f(t)$ is log-symmetric (i.e., $f(t) = f(1/t)$). When $t = 2$, $f(t) = 9/10$ and when $t=3$, $f(t) = 4/5$. That is to say, with $t = 2$, we can shrinkage CI width by roughly 5\% and with $t = 3$, we can shrinkage CI width by 11\%.

\begin{figure}[H] 
    \centering
    \includegraphics[width=0.6\columnwidth]{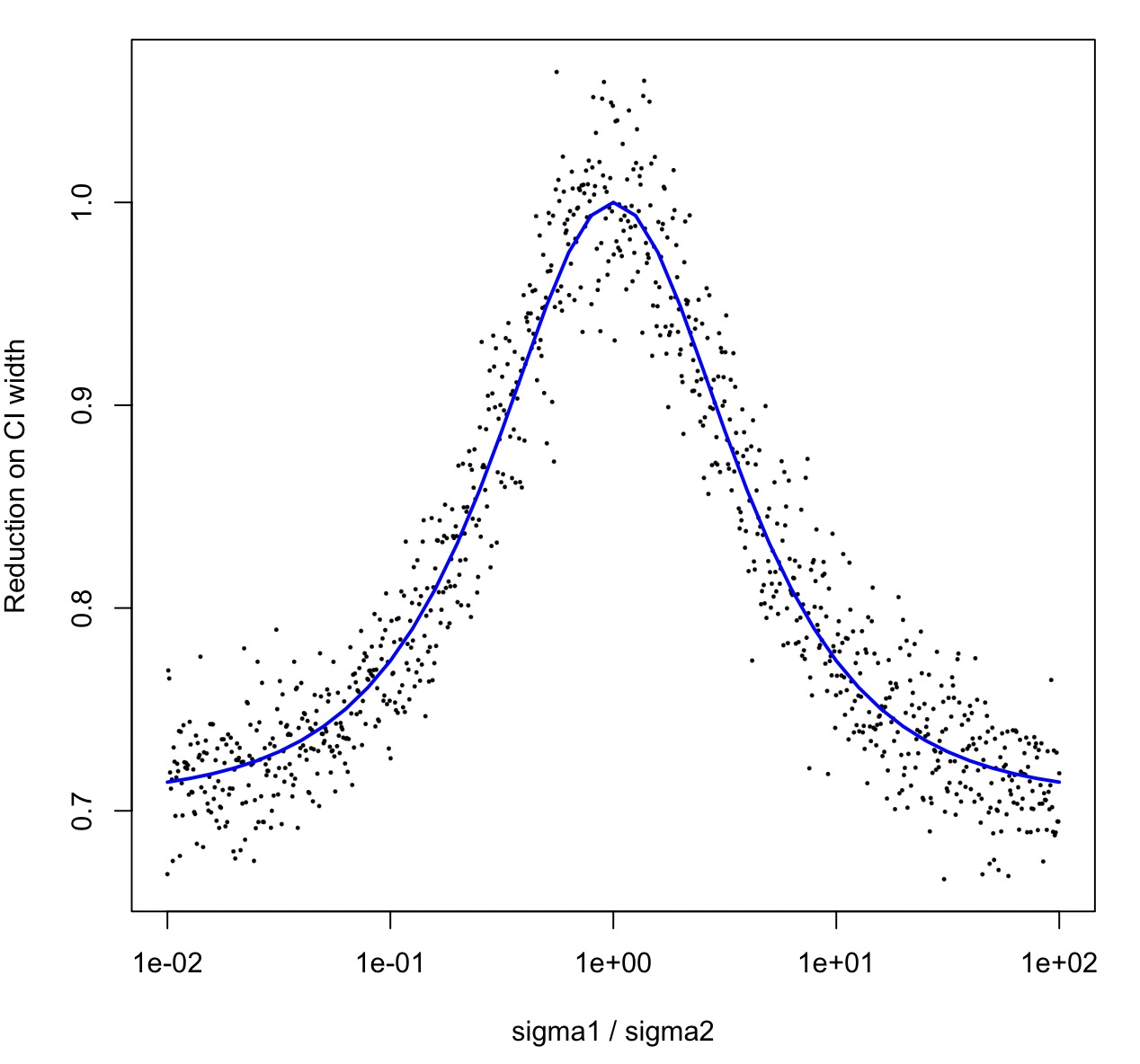}
    \caption{\textbf{Simulation on Confidence interval reduction:} Simulation study showing the relationship between reduction on CI against the ratio between $\sigma_1$ and $\sigma_2$ on a two-group simulated top example where $N_1 = N_2 = 10000$ and total sampling budget $n = 1000$. Blue line shows the theoretical function. }
    \label{demo}
\end{figure}

\begin{figure}[H] 
    \centering
    \includegraphics[width=1\columnwidth]{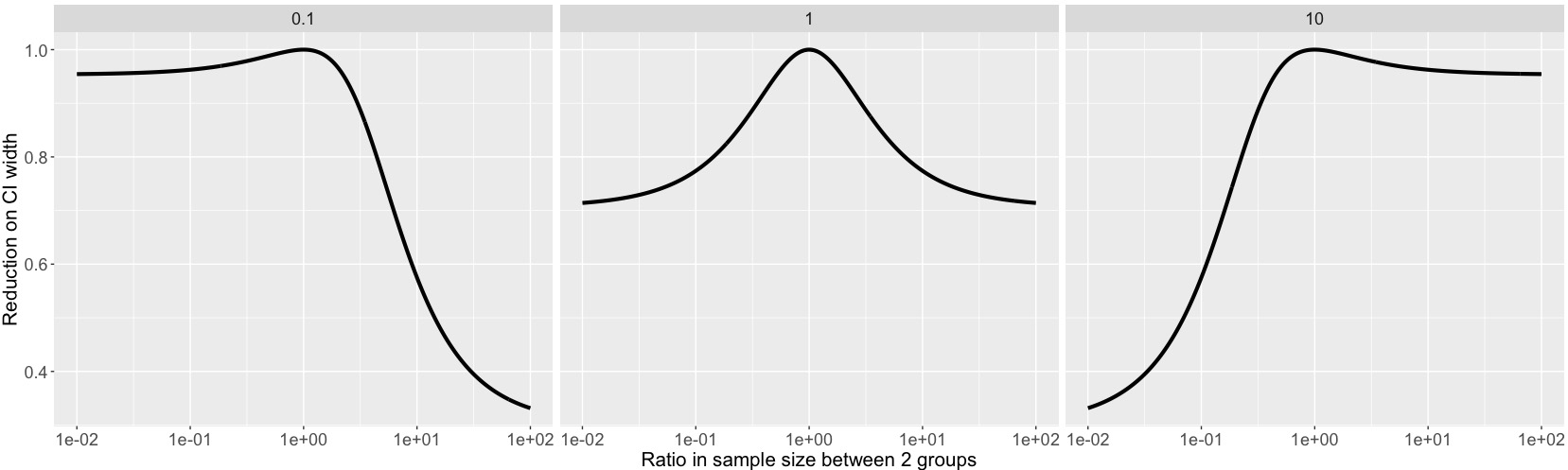}
    \caption{\textbf{Confidence interval reduction by $\sigma_1 / \sigma_2$ across different $N_1/N_2$:} Relationship between reduction on CI against the ratio between $\sigma_1$ and $\sigma_2$ on a two-group simulated top example where $N_2/N_1$ being 0.1, 1 and 10.}
    \label{demo}
\end{figure}

We ran a simulation example under the simple 2-group setting: $N_1 = N_2 = 10000$ but the sampling budget is only $n = 1000$ (which is only 5\% of the total sample). With a fixed $\sigma_1 = 1$, we simulated $\sigma_2$ from 0.01 to 100 and evaluated the variance of the estimation using 1000 parallel computations.

The above simulation setting assumes equal sample size $N_1 = N_2$. It can be generalized with one additional hyperparameter $N_2 / N_1$. In the simulation study discussed in Figure 2, the expected variance reduction is a function of $N_2/N_1$ and $\sigma_2 / \sigma_1$ following $$\frac{V_*}{V_0} = \frac{\left(1+N_2/N_1 \sigma_2 \ \sigma_1 \right)^2}{(1 + N_2/N_1 (\sigma_2 / \sigma_1)^2) (1+N_2/N_1)}$$

The above discussion assumes the two group setting where we can achieve the variance reduction substantially. When we increase the number of groups (assuming equal size), the expected variance reduction follows

$$\frac{\left(1 + \left(\frac{\sigma_2}{\sigma_1}\right) + ... + \left(\frac{\sigma_k}{\sigma_1}\right)\right)^2}{k \left(1+\left(\frac{\sigma_2}{\sigma_1}\right)^2 + ... + \left(\frac{\sigma_k}{\sigma_1}\right)^2 \right)}$$

In this simulation example, we simulated from $k = 1$ and $k = 200$ and under each simulation setting, the ratio between $\sigma_k$ and $\sigma_1$ was simulated following $10^Y$ where $Y$ follows normal distribution with mean 0 and variance 2. From the figure showing here, the benefit on \verb+BayesSRW+ keeps increasing with larger number of groups, although the most immediate effect happens when $k$ is small. This is as expected as two folds:

\begin{itemize}
    \item The higher number of groups, the more information regarding $\sigma_k / \sigma_1$ will be taken into account for \verb+BayesSRW+ compared to the uniform sampling approach. 
    \item The Bayes approach, taking the distribution $p(\mu, \sigma)$ for the data inference, will get more benefits with high dimensional or large scale settings \citep{zheng2021mixtwice, zheng2024bootstrap}. 
\end{itemize}

\begin{figure}[H] 
    \centering
    \includegraphics[width=0.7\columnwidth]{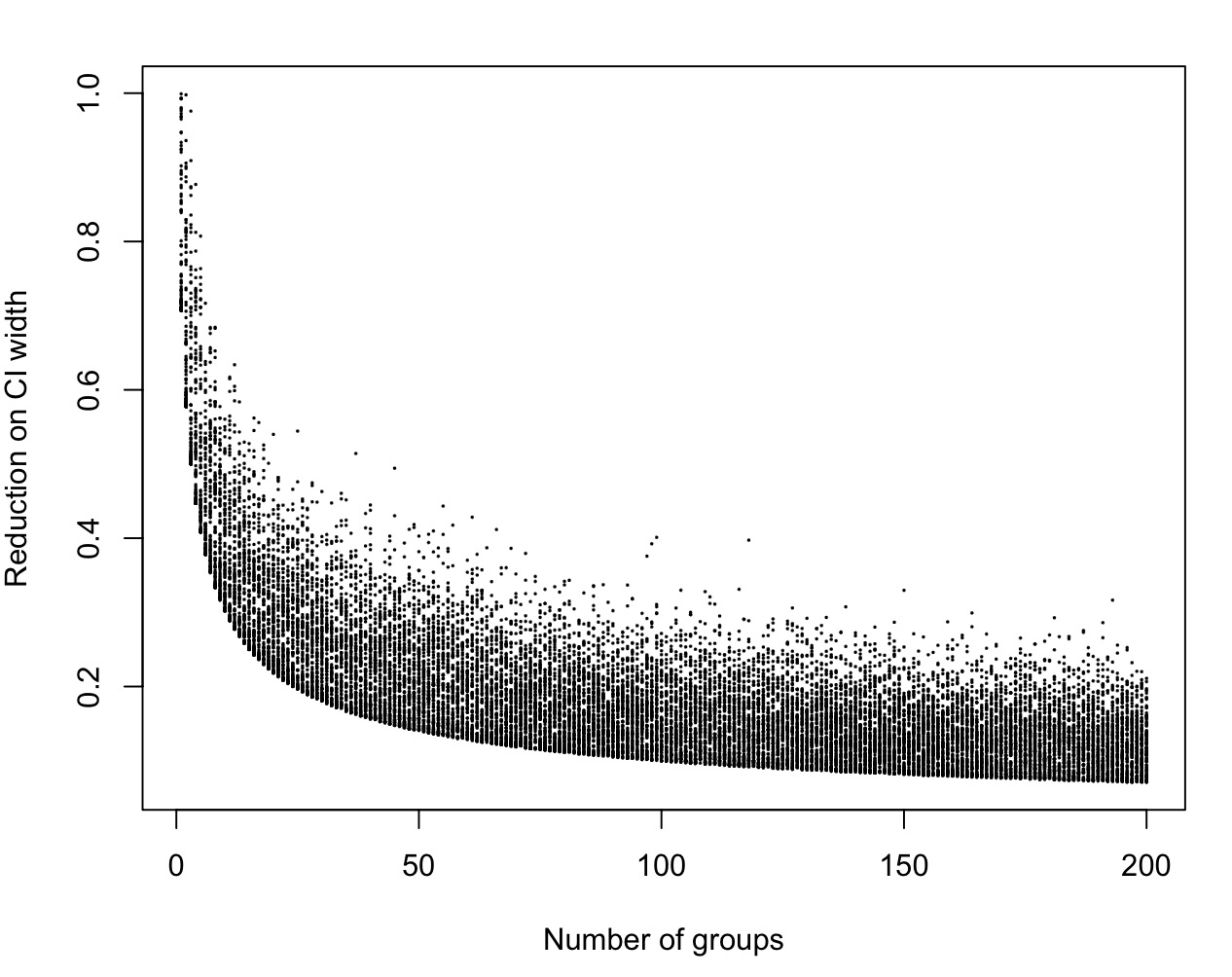}
    \caption{\textbf{Confidence interval reduction by number of groups $k$:} Simulation example showing the relationship between CI reduction against number of groups $k$ going from 1 to 200.}
    \label{demo}
\end{figure}

\begin{figure}[H] 
    \centering
    \includegraphics[width=1\columnwidth]{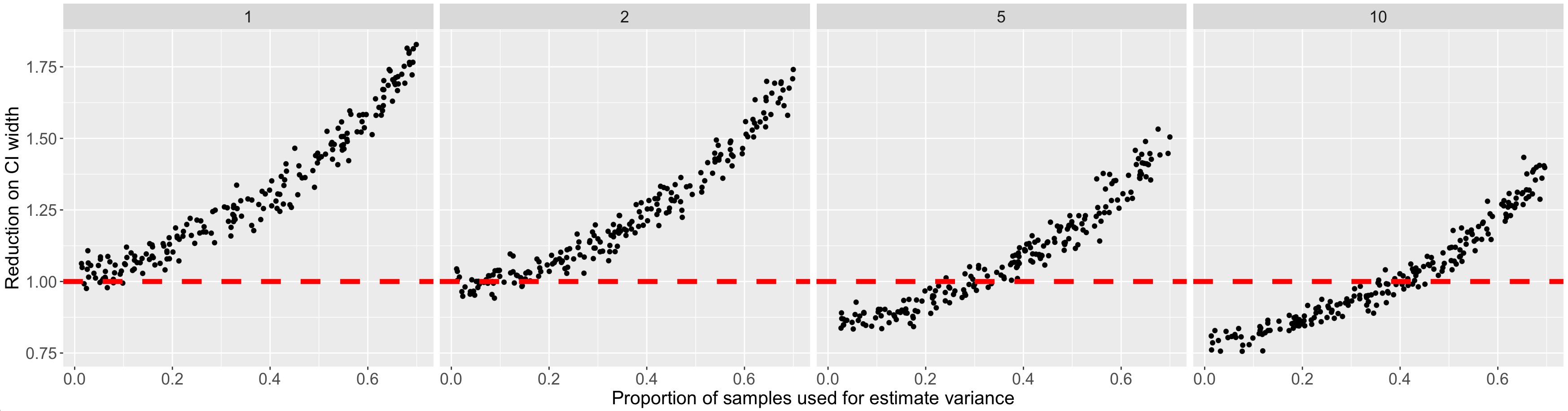}
    \caption{\textbf{How many samples should be used to estimate variance and fit the prior distribution:} Simulation example showing the relationship between CI reduction against proportion of samples $p$ to be used to estimate prior (especially variance distribution) under the settings with $\sigma_2 / \sigma_1$ being 1, 2, 5, 10 etc.}
    \label{demo}
\end{figure}

For practical guidance, there is one additional hyperparameter that needs specifically input for the algorithm, which is the proportion of samples ($p$) that can be used for the prior knowledge, particularly to estimate the variance ratio $\sigma_2 / \sigma_1$ or the ratios $\sigma_k / \sigma_1$ with high dimensional settings. Suppose the sampling budget is $n$, the more samples ($np$) delegated for variance estimation the fewer samples ($n(1-p)$) can be left for updating the prior therefore for the real parameter $\mu$ estimation. For simplifying the discussion, a simulation example was studied with the same setting as in Figure 1 but under the scenarios of $\sigma_2/\sigma_1 = 1, 2, 5, 10$. It's expected that the higher number used for variance estimation, the higher CI we will get for estimating $\mu$. Also, consistent with the intuition above, the higher the variance ratio, the more benefit on CI reduction we can get by running \verb+BayesSRW+. Under the degenerated scenario where $\sigma_1 = \sigma_2$, we should not devote any samples for the variance estimation. Under the case where $\sigma_2 = 10\sigma_1$, we should still outperform the baseline approach even though half of the samples are used to estimate the variance. From this study, the general practical guidance should be use minimum proportion of samples out of the budget to run variance estimation and to have a baseline prior $p(\mu, \sigma)$ and use the majority of the samples to update and get posterior estimation. E.g., when you have sampling budget $n = 1000$, it makes sense to only use 10\% or even 5\% of the samples. More samples can be used in the high dimensional settings with multiple groups. This aligns with the intuition of most of the Bayes study as the final M.A.P estimation is very robust against the variability introduced in prior.

\section{Conclusion and Future Directions}

In this work, we have explored the challenges and strategies involved in optimal survey sampling and estimation under constraints, particularly focusing on scenarios where variance across different groups plays a crucial role. Through a systematic approach grounded in Cauchy inequality, we demonstrated that the conventional method of uniform sampling, while providing unbiased estimates, often leads to suboptimal results due to higher variance. Instead, our proposed methodology, \verb+BayesSRW+, which allocates samples proportionally to both group size and variance in a Bayes framework, offers a significant improvement in terms of minimizing estimation variance.

The proposed two-stage sampling and re-weighting algorithm addresses practical limitations by incorporating a preliminary phase dedicated to estimating the unknown variance parameters. This approach is particularly relevant in real-world applications like the survey framework, where prior knowledge of variance is often limited, yet essential for efficient resource allocation. By strategically dividing the sampling budget between variance estimation and optimal sampling, our method ensures a more precise and reliable estimation of the population mean, even when initial information is scarce.

The work presented here opens several avenues for future research. One immediate extension is the exploration of adaptive sampling methods that dynamically adjust the allocation of resources as more data becomes available. Such methods could further enhance the efficiency of the sampling process, particularly in online or sequential settings.

Another promising direction involves the application of this methodology to more complex models, including those with hierarchical or multilevel structures. In these cases, the interplay between variance at different levels of the hierarchy could be explored to refine the optimal allocation strategy further.

Moreover, the integration of machine learning techniques to estimate variance components in high-dimensional or sparse data scenarios could provide valuable insights and enhance the practical applicability of the proposed methods. This would be particularly relevant in settings with large-scale or real-time data, where traditional variance estimation techniques may fall short.

Finally, while this study focused on minimizing variance in the context of population mean estimation, similar principles could be applied to other statistical metrics, such as quantiles or regression coefficients. Extending the framework to these settings could yield novel strategies for optimal sampling and resource allocation, broadening the impact and utility of the approach.

\clearpage
\bibliography{references}

\end{document}